\documentclass[5p,twocolumn,times]{elsarticle}
\usepackage{lipsum}
\usepackage{lineno,hyperref}
\modulolinenumbers[5]
\usepackage[pdftex]{color}
\usepackage[font=footnotesize,labelfont=bf]{caption}
\usepackage[font=footnotesize,labelfont=bf]{subcaption}
\journal{Journal of \LaTeX\ Templates}
\usepackage{colortbl}
\usepackage{xcolor}

\usepackage{amssymb}

\biboptions{compress}

\usepackage[figuresright]{rotating}

\begin{document}

\begin{frontmatter}


\title{Locally adaptive aggregation of organisms under death risk in rock-paper-scissors models}


\address[1]{Institute for Biodiversity and Ecosystem
Dynamics, University of Amsterdam, Science Park 904, 1098 XH
Amsterdam, The Netherlands}
\address[2]{School of Science and Technology, Federal University of Rio Grande do Norte\\
Caixa Postal 1524, 59072-970, Natal, RN, Brazil}
\address[3]{Department of Computer Engineering and Automation, Federal University of Rio Grande do Norte, Av. Senador Salgado Filho 300, Natal, 59078-970, Brazil}

\author[1,2]{J. Menezes}  
\author[2]{E. Rangel} 

\begin{abstract}
We run stochastic simulations of the spatial version of the rock-paper-scissors game, considering that individuals use sensory abilities to scan the environment to detect the presence of enemies. If the local dangerousness level is above a tolerable threshold, individuals aggregate instead of moving randomly on the lattice.
We study the impact of the locally adaptive aggregation on the organisms' spatial organisation by measuring the characteristic length scale of the spatial domains occupied by organisms of a single species. 
Our results reveal that aggregation is beneficial if triggered when the local density of opponents does not exceed $30\%$; otherwise, the behavioural strategy may harm individuals by increasing the average death risk. We show that if organisms can perceive further distances, they can accurately scan and interpret the signals from the neighbourhood, maximising the effects of the locally adaptive aggregation on the death risk. 
Finally, we show that the locally adaptive aggregation behaviour promotes biodiversity independently of the organism's mobility. The coexistence probability rises if organisms join conspecifics, even in the presence of a small number of enemies. We verify that our conclusions hold for more complex systems by simulating the generalised rock-paper-scissors models with 
five and seven species.
Our discoveries may be helpful to ecologists in understanding 
systems where organisms' self-defence behaviour adapts to local environmental cues.
\end{abstract}


\end{frontmatter}



\section{Introduction}

Behavioural biology has revealed the mechanisms that organisms 
use to improve their fitness, being fundamental for the stability
of the rich biodiversity in nature\cite{ecology,Nature-bio,BUCHHOLZ2007401,doi:10.1098/rstb.2019.0012}. 
There is plenty of evidence that self-preservation strategies are properly executed because of the organism's evolutionary ability to scan the environment cues, perceiving the presence of a nearby enemy and the energy expended in the action\cite{Cost2,olfactory2,LizardB1,detection,adaptive2}. In this scenario, 
living in groups facilitates the defence action since individual protection against enemies is maximised by collective effort in surveillance and resistance, demanding less individual energy expenditure on defense against enemies \cite{manyeyes,dilution1, dilution2,Grouping1,Grouping2,fishing,strategy1,strategy4,strategy5,strategy3}. 

Cyclic models of biodiversity have been studied using the rock-paper-scissors game rules, which successfully describe the nonhierarchical competition interactions found in many biological systems \cite{Avelino-PRE-86-036112,uneven,Moura, EXPANDING, Bazeia_2017, Anti1,anti2,eloi,Menezes_2022,TENORIO2022112430}. However, experiments with bacteria \textit{Escherichia coli} revealed
that the cyclic dominance among three bacteria strains is insufficient to stabilise the system. It has been discovered that coexistence is ensured only if individuals interact locally \cite{Allelopathy}. This shows the central role of space in the stability of biological systems, as it has been also observed in communities of lizards and systems of competing coral reefs \cite{lizards,coral,Extra1}. Furthermore, cyclic dominance has been shown to play a fundamental role in the spatial interactions in social systems, public good with punishment, and human bargaining \cite{Rev2,Rev3}.

There is plenty of evidence that organisms' mobility plays a central role in promoting or jeopardising biodiversity in structured populations
\cite{tanimoto2,Rev1,Rev6,mobilia2,weakest,doi:10.1021/ja01453a010,Volterra,PhysRevE.78.031906,PhysRevE.82.066211,PARK10,PARK11}. 
Evidence shows that organisms' foraging behaviour may affect biodiversity in the spatial rock-paper-scissors game \cite{Moura,Menezes_2022}. Organisms' moving to escape their enemies and find natural resources to the species perpetuation may unbalance the cyclic game or decelerate the population dynamics, thus jeopardising or promoting biodiversity \cite{Menezes_2022,TENORIO2022112430,Adaptive1,Rev4,MENEZES2022101606}. 

Recently, it has been shown that aggregation behaviour is an efficient antipredator strategy in tritrophic predator-prey cyclic models \cite{MENEZES2022101606}. Numerical simulations of the Lotka-Volterra version of the rock-paper-scissors game revealed that individuals' predation risk decreases if organisms execute a gregarious movement, instead of exploring the territory to found prey and reproduce.
In contrast with the standard model where organisms move in a random direction, the grouping strategy produces spiral-type patterns 
with organisms of the same species living in spatial domains whose characteristic length depends on the 
 the distance the individuals can scan their neighbourhood, and their cognitive ability to perform the directional self-preservation movement tactic \cite{MENEZES2022101606}. 
 
Although the revealing details of the complexity of the spatial interactions, the model in Ref.~\cite{MENEZES2022101606} considers exclusively a non-adaptive aggregation tactic, i.e., individuals cannot smartly adapt their movement to trigger the grouping strategy only when pressured by an imminent enemy' attack, as happens, for example, in spider mites communities \cite{adaptive2}. 
In this case, the unnecessary expenditure is avoided since organisms can continue freely advancing on the lattice to conquer territory, allowing the population growth \cite{mobilia2}.
In this work, we sophisticate the stochastic model to simulate a locally adaptive aggregation where organisms move gregariously only under death risk \cite{MENEZES2022101606}. We also consider that the decision to aggregate is the individual competence, meaning that each organism acts autonomously according to its own local reality. Therefore, each individual can decide if moving gregariously or randomly, with the congregation being triggered only if the local density of enemies is higher than a tolerable threshold. In addition, we implement the behavioural survival strategy using the May-Leonard implementation of the spatial rock-paper-scissors game. This allows the generalisation of our results to systems where competition for natural resources is the goal of the cyclic game \cite{leonard}. 

We aim to answer the following questions: i)
how does the locally adaptive aggregation modify the spiral patterns, characteristic of the standard May-Leonard implementation of the rock-paper-scissors model?;  
ii)
how does the aggregation trigger influence the organisms' spatial organisation altering the size of the typical single-species domains?;
iii)
how does adaptive grouping benefit individuals by reducing the average death risk?;
iv)
how does the locally adaptive congregation behaviour impact species coexistence probability?

\begin{figure}
\centering
\includegraphics[width=40mm]{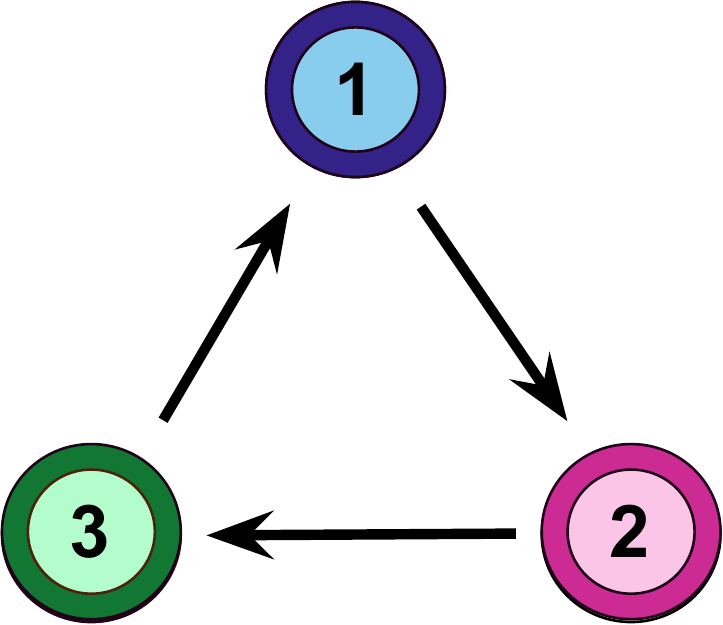}
\caption{The rock-paper-scissors model rules. The black arrows illustrate the dominance in the spatial game: individuals of species $i$ eliminate organisms of species $i+1$, with $i=1,2,3$ and $i\pm3 = i$. Organisms of the same species aggregate when attacked and move randomly when not in danger. Dark blue, pink, and green represent individuals of species $1$, $2$, and $3$ moving gregariously; light blue, pink, and green indicate organisms of species $1$, $2$, and $3$ moving randomly.}
	\label{fig1}
\end{figure}
\begin{figure*}[h]
\centering
    \begin{subfigure}{.23\textwidth}
        \centering
        \includegraphics[width=40mm]{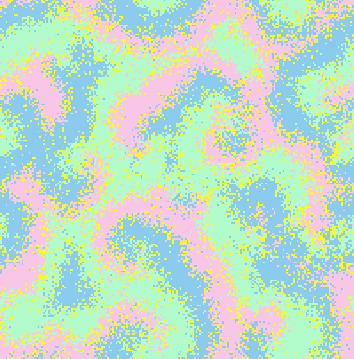}
        \caption{}\label{fig2a}
    \end{subfigure} %
       \begin{subfigure}{.23\textwidth}
        \centering
        \includegraphics[width=40mm]{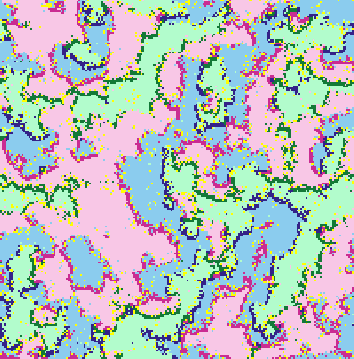}
        \caption{}\label{fig2b}
    \end{subfigure} %
   \begin{subfigure}{.23\textwidth}
        \centering
        \includegraphics[width=40mm]{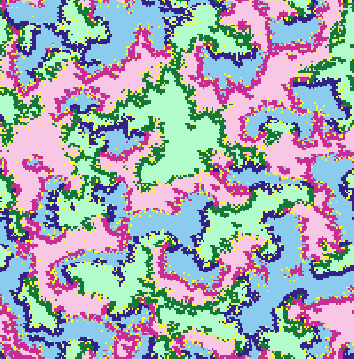}
        \caption{}\label{fig2c}
    \end{subfigure} 
   \begin{subfigure}{.23\textwidth}
        \centering
        \includegraphics[width=40mm]{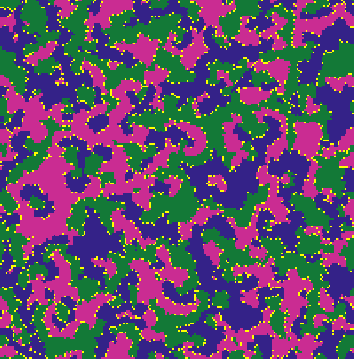}
        \caption{}\label{fig2d}
            \end{subfigure} 
\caption{Snapshots captured from simulations of the rock-paper-scissors game with individuals' locally adaptive aggregation. The realisations ran in lattice with $200^2$ grid points for a timespan of $2000$ generations, with $R=3$, $r=s=0.25$ and $m=0.5$.
Figures \ref{fig2a}, \ref{fig2b}, \ref{fig2c}, and \ref{fig2d} show the organisms' spatial organisation at the end of Simulation A ($\varphi=1.0$), B ($\varphi=0.1$), ($\varphi=0.025$), and D ($\varphi=0.0$), respectively. The colours follow the scheme in Fig. \ref{fig1}, with blue, pink, and green depicting individuals of species $1$, $2$, and $3$, respectively. Dark and light colours distinguish organisms performing the congregation strategy and moving randomly. Yellow dots depict empty sites. 
} 
\label{fig2}
\end{figure*}
\begin{figure*}
\centering
       \begin{subfigure}{.48\textwidth}
        \centering
        \includegraphics[width=85mm]{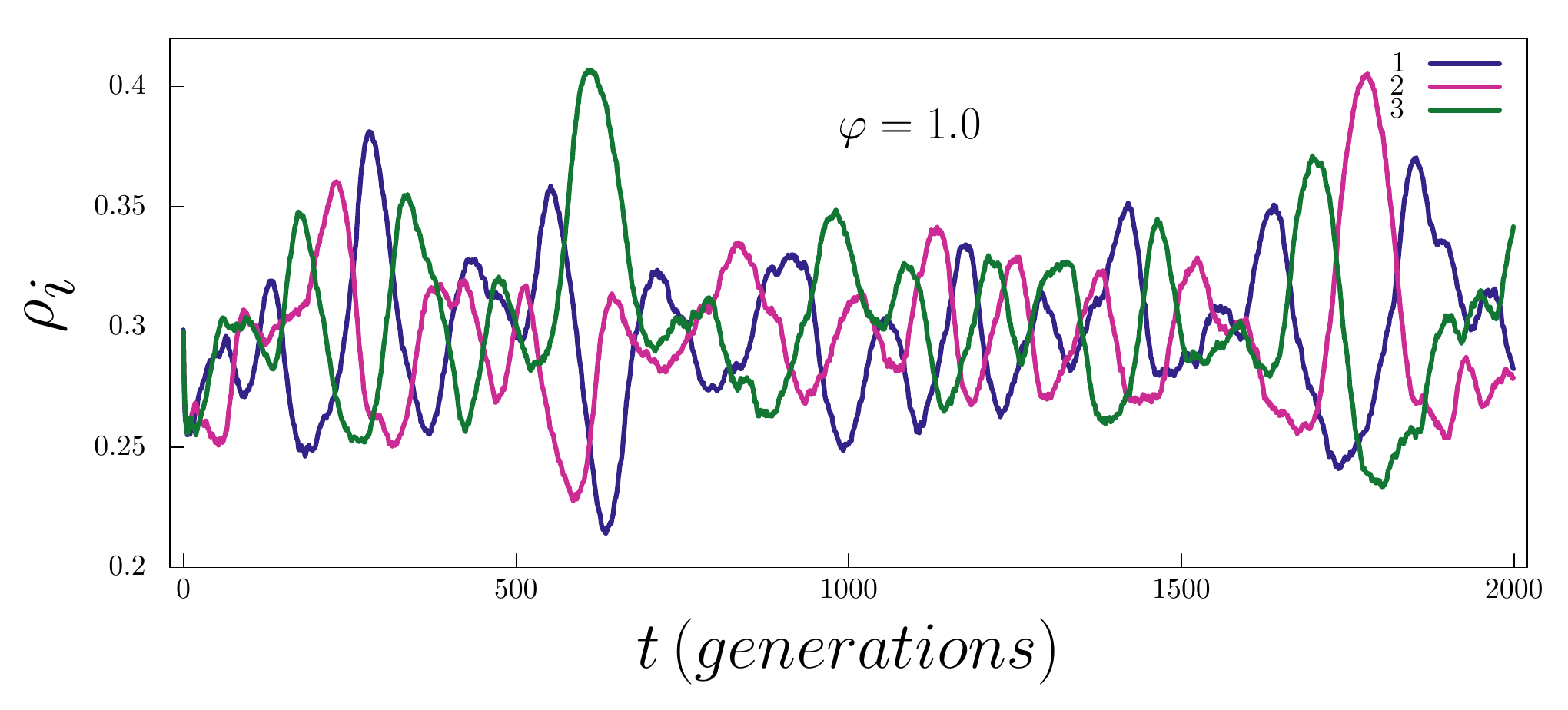}
        \caption{}\label{fig3a}
    \end{subfigure}
           \begin{subfigure}{.48\textwidth}
        \centering
        \includegraphics[width=85mm]{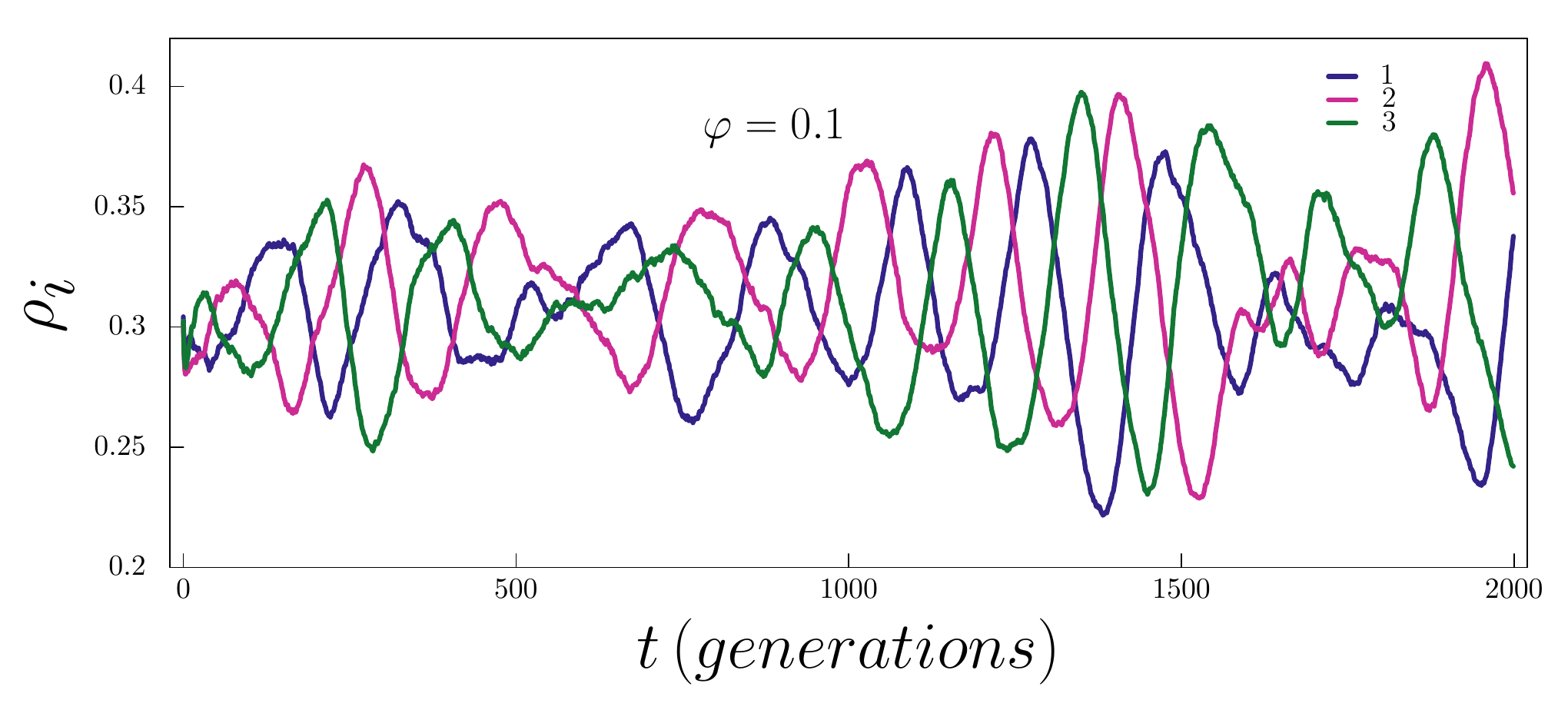}
        \caption{}\label{fig3b}
    \end{subfigure}\\
       \begin{subfigure}{.48\textwidth}
        \centering
        \includegraphics[width=85mm]{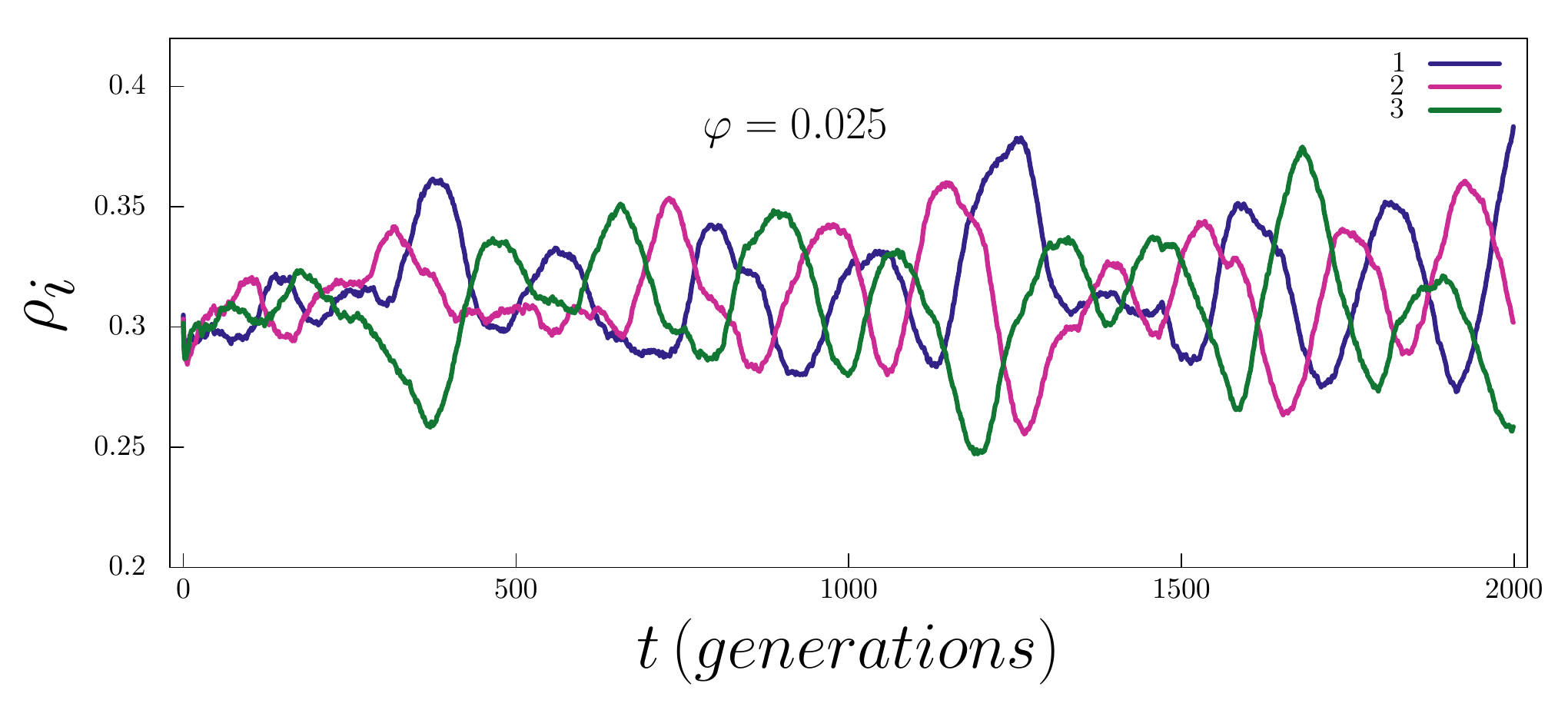}
        \caption{}\label{fig3c}
    \end{subfigure}
           \begin{subfigure}{.48\textwidth}
        \centering
        \includegraphics[width=85mm]{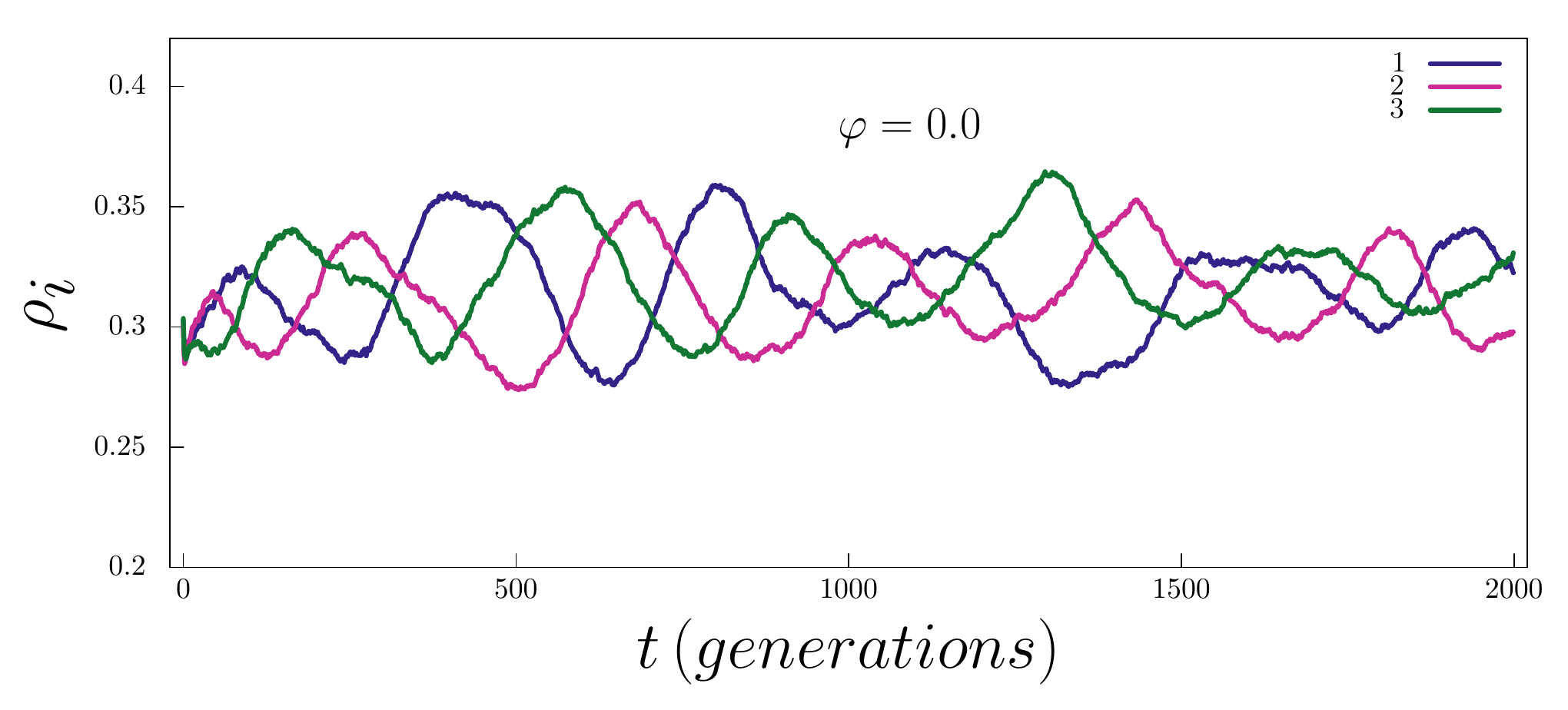}
        \caption{}\label{fig3d}
    \end{subfigure}
\caption{Dynamics of species densities during the simulations in Fig.~\ref{fig2}. The blue, pink, and green lines in Figs.~\ref{fig3a}, 
~\ref{fig3b}, ~\ref{fig3c}, and ~\ref{fig3d} depict the temporal dependence of the density of individuals of species $1$, $2$, and $3$, in Simulations A, B, C, and D, respectively.
} 
\label{fig3}
\end{figure*}
\begin{figure}
	\centering
	\includegraphics*[width=8.5cm]{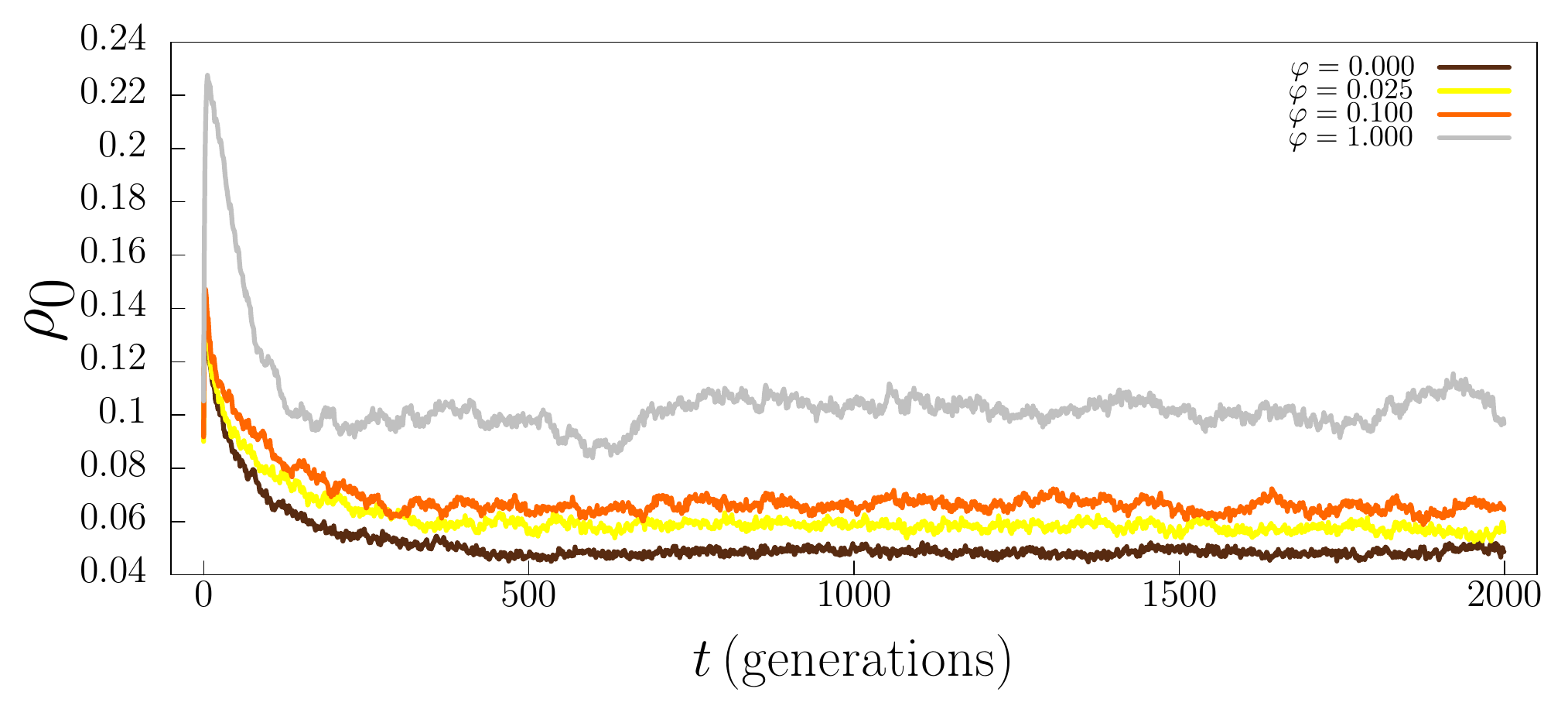}
\caption{Temporal dependence of the density of empty spaces in simulations of Fig.~\ref{fig2}. The grey, orange, yellow, and brown lines show the dynamics of empty sites in Simulations A, B, C, and D, respectively.}
 \label{fig4}
\end{figure}


The outline of this paper is as follows. In Sec.~\ref{sec2}, we introduce our stochastic model and present the methods used to implement the locally adaptive grouping in our simulation algorithm. In Sec. \ref{sec3}, the changes in the spatial patterns are studied for various 
values of aggregation trigger; the autocorrelation function and characteristic length scales are addressed in Sec. \ref{sec4}. 
The reduction in the organisms' average death risk is computed in Sec. \ref{sec5} for a range of aggregation triggers and mobility probabilities. Finally, the coexistence probability in terms of the individual's mobility is investigated in Sec.~\ref{sec6}, while our comments and conclusions appear in Sec.~\ref{sec7}.

\label{Introduction}

\section{The Model}
\label{sec2}

We study a cyclic model of three species that outcompete each other according to the rock-paper-scissors game rules, illustrated in Fig.~\ref{fig1}. This means that individuals of species $i$ eliminate organisms of species $i+1$, with $i=1,2,3$, with the cyclic identification $i=i+3\,\beta$, where $\beta$ is an integer. 
Our model considers that organisms of the same species aggregate to minimize the probability of being killed in the spatial game. The gregarious movement is locally adaptive, 
triggered whenever the density of enemies in the organisms' neighbourhood is higher than a tolerable threshold. This means that each individual of species $i$ can scan the environment to perceive the presence of organisms of species $i+1$, thus, accurately deciding if the best strategy is to search for refuge joining their conspecifics.
or continue moving randomly to explore the territory. The dark colours in Fg.\ref{fig1} stand for individuals executing the gregarious movement, whereas the light colours represent organisms moving randomly.

\subsection{Numerical simulations}

To perform the numerical simulations, we use square lattices with periodic boundary conditions; the number of grid sites is $\mathcal{N}$. We use the May-Leonard implementation, where the total number of individuals is not conserved. Therefore, as each grid point is occupied by at most one individual (or it is empty), the maximum number of organisms in the system is the total number of grid points $\mathcal{N}$.

Initially, the organisms are randomly distributed in the lattice: each individual is allocated at a random grid site. The initial conditions are prepared so that the number of individuals is the same for every species is the same. We define the number of individuals of each species at the initial state as one-third of the total number of organisms: $I_i (t=0)\,\approx \,\mathcal{N}/3$, with $i=1,2,3$; the rest of grid sites are left empty in the initial conditions.

Once the random initial conditions are ready, the algorithm stochastically implements the interactions following the von Neumann neighbourhood, where each organism can interact with one of its four immediate neighbours.
The spatial interactions are:
\begin{itemize}
\item
Selection: $ i\ j \to i\ \otimes\,$, with $ j = i+1$, where $\otimes$ means an empty space: an individual of species $i$ eliminates a neighbour of species $i+1$ following the rules illustrated in Fig.\ref{fig1} - the grid site occupied by the eliminated individual is left empty.
\item
Reproduction: $ i\ \otimes \to i\ i\,$: an empty space is filled by a new organism of any species.
\item
Mobility: $ i\ \odot \to \odot\ i\,$, where $\odot$ means either an individual of any species or an empty site. An organism moves by switching positions with another individual of any species or an empty space.
\end{itemize}

The interactions are implemented following a fixed set of probabilities which is the same for every species: $s$ (selection probability), $r$ (reproduction probability), and $m$ (mobility probability). During the interaction implementation, the code follows the steps:

\begin{enumerate}
\item
an active individual of any species is drawn among all organisms in the lattice;
\item
one interaction is randomly chosen following the set of probabilities rates ($s$, $r$, and $m$);
\item
one of the four immediate neighbours is drawn to suffer the action (selection, reproduction, and random mobility) - the only exception is the adaptive gregarious movement, where the organism move towards the direction with more conspecifics.
\end{enumerate}
Every time an interaction is implemented, one timestep is counted. After $\mathcal{N}$ timesteps, one generation is completed - our time unit is one generation.

To understand the population dynamics during the simulations, we calculate the density of organisms of species $i$, $\rho_i(t)$, with $i=1,2,3$. This is defined as the fraction of the lattice occupied by individuals of the species $i$ at time $t$, $\rho_i(t)=I_i(t)/\mathcal{N}$. Also, the temporal dependence of the density of empty spaces is computed as $\rho_0 = 1 - \rho_1 - \rho_2 - \rho_3$.

\subsection{Implementing the locally adaptive aggregation strategy}

To implement the locally adaptive grouping tactic, we define the perception radius, $\mathcal{R}$, to represent the maximum distance an organism of species $i$ can scan the environment to be aware of the presence of enemies. Thus, the local density of organisms of each species is computed within a circular area of radius $\mathcal{R}$, centred in the organism of species $i$ \cite{TENORIO2022112430,MENEZES2022101606}.
In addition, we introduce the aggregation trigger, $\varphi$, to represent the minimum local density of individuals of species $i-1$ (enemies) that stimulates the organism of species $i$ to move gregariously. This means that if the local density of organisms of species $i-1$ is lower than $\varphi$, the individual moves randomly.

The numerical implementation of the gregarious movement is performed by dividing the observing disc into four circular sectors, each section in the directions of the one nearest neighbour of the von Neumann neighbourhood \cite{Moura,Anti1,anti2,Menezes_2022,MENEZES2022101606,combination}. Next, it is computed how many individuals of species $i$ exist within each circular sector, with organisms on the circular sector borders assumed to be part of both circular sectors. Finally, the organism switches positions
with the immediate neighbour in the direction with more conspecifics; a draw in the event of a tie.


\section{Spatial Patterns}
\label{sec3}

Our first goal is to understand the effects of the locally adaptive 
congregation strategy in spatial patterns. Therefore, we ran
a single simulation for four values of the aggregation trigger: 
\begin{itemize}
\item
Simulation A: $\varphi=1.0$ - the absence of organisms' grouping behaviour, i.e., individuals do not aggregate even under death risk;
\item
Simulation B: $\varphi=0.1$ - organisms' agglomeration occurs if, at least, $10\%$ neighbours are enemies;
\item
Simulation C: $\varphi=0.025$ - an individual move gregariously if 
at least, $2.5\%$ neighbours are enemies;
\item
Simulation D: $\varphi=0.0$ - the gregarious movement is not locally adaptive, with individuals always grouping independently of the presence of enemies surrounding them.
\end{itemize}
The realisations were performed in lattices with $200^2$ grid sites, running for a timespan of $2000$ generations. We set the parameters to $s=r=0.25$, $m=0.5$, and $R=3$.

Figures \ref{fig2a}, \ref{fig2b}, \ref{fig2c}, and \ref{fig2d} show the individuals' spatial organisation at the end of Simulations A, B. C, and D, respectively. 
To depict each organism, we use the same colours of the scheme in Fig.~\ref{fig1}: blue, pink, and green dots show the individuals of species $1$, $2$, and $3$, respectively. The organisms performing the aggregation strategy are highlighted using dark colours, while the individuals moving randomly appear in light shades.
We also quantified the dynamics of the species densities for Simulation A, B, C, and D, which are depicted in Figs.~\ref{fig3a}, \ref{fig3b}, \ref{fig3c}, and \ref{fig3d}, respectively. As in Fig.\ref{fig1}, blue, pink, and green lines shows the temporal dependence of densities of individuals of species $1$, $2$, and $3$, respectively;

Let us first focus on Simulation A, where individuals do not aggregate to protect themselves against enemies (Fig.~\ref{fig2a}). Because of the random initial conditions, selection interactions are frequent at the beginning of the simulation. After that, spatial patterns are formed with organisms of the same species occupying departed patches. Since organisms are unaware of the neighbourhood, they move randomly, independently of the risk of being caught. This results in faster dynamics of species densities, with organisms being destroyed and newborns appearing at a high rate. Consequently, the species densities' frequency and amplitude are high, as shown in Fig.~\ref{fig3a}.

In addition to the usual pattern formation process driven by the cyclic game rules, the gregarious movement performed by individuals under death risk promotes the formation of self-protection clusters on the border that is attacked by enemies, as shown in Figs.~\ref{fig2b} and ~\ref{fig2c}. For example, the organisms of species $2$ aggregating (dark pink dots) are concentrated on the border with spatial domains of species $1$ (blue areas). The self-preservation movement tactic produces a deformation of the spiral patterns, with individuals concentrating in patches with smaller sizes since they abdicate to explore extensive areas of the territory to form clumps. Because of this, the population dynamics are decelerated, with reduced frequency and amplitude, as depicted in Figs.\ref{fig3b} and \ref{fig3c}.

Finally, the snapshot in Fig.~\ref{fig3d} reveals what occurs in the case of the non-adaptive aggregation strategy ($\varphi=0.0$) - individuals move gregariously even if no enemy surrounds them. In this scenario, the population dynamics are altered since the individuals neglect the conquest of new territories 
to focus exclusively on the survival movement strategy. This induces a contraction of the spatial domains occupied by organisms of a single species, since individuals do not advance in the territory even if they are not under death risk.
Finally, Fig.~\ref{fig4} shows the temporal dependence of the density of empty spaces, $\rho_0$, in Simulations A (grey line), B (orange line), C (green line), and D (brown line). The results show that the density of empty spaces decreases after
an initial period of pattern formation. Furthermore, the locally congregation reduces the organisms' death risk. Because of this, the lower the aggregation trigger, the more the density of empty spaces is reduced.

\begin{figure}[h]
	\centering
	\includegraphics*[width=8.5cm]{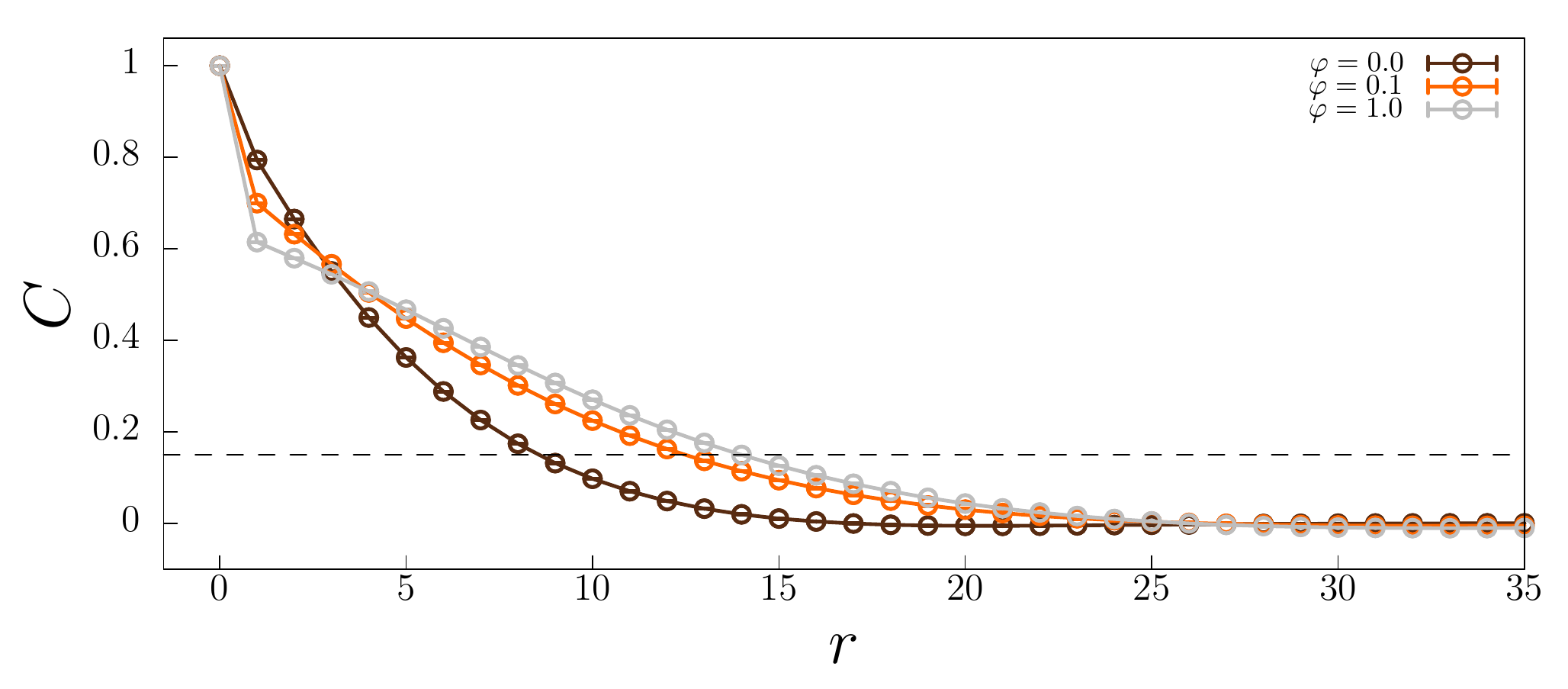}
\caption{Autocorrelation functions in terms of the radial coordinate. The grey, orange, and brown lines depict the results for the standard model ($\varphi=1.0$), aggregation triggered when at least $10\%$ of neighbours are enemies
$\varphi=0.1$, and the non-adaptive aggregation ($\varphi=0.0$), respectively. The error bars indicate the standard deviation; the dashed black line shows the threshold assumed to calculate the characteristic length scale. The interaction probabilities are $r=s=0.25$ and $m=0.5$; the perception radius is $R=3$.
} 
 \label{fig5}
\end{figure}

\section{Autocorrelation Function}
\label{sec4}

Let us now quantify the scale of spatial domains in the presence of locally adaptive aggregation. For this, we compute the spatial autocorrelation function.
The autocorrelation function is computed from the inverse Fourier transform of
the spectral density as
\begin{equation}
C(\vec{r}') = \frac{\mathcal{F}^{-1}\{S(\vec{k})\}}{C(0)},
\end{equation}
where $S(\vec{k})$ is given by
\begin{equation}
S(\vec{k}) = \sum_{k_x, k_y}\,\Phi(\vec{\kappa}),
\end{equation}
and $\Phi(\vec{\kappa})$ is Fourier transform
\begin{equation}
\Phi(\vec{\kappa}) = \mathcal{F}\,\{\phi(\vec{r})-\langle\phi\rangle\}.
\end{equation} 
The function $\phi(\vec{r})$ represents the spatial distribution of individuals of species $1$, with $\phi(\vec{r})=0$ and $\phi(\vec{r})=1$ indicating the absence and the presence of an individual of species $1$ in at the position $ \vec{r}$ in the lattice, respectively). The spatial autocorrelation function is given by
\begin{equation}
C(r') = \sum_{|\vec{r}'|=x+y} \frac{C(\vec{r}')}{min \left[2N-(x+y+1), (x+y+1)\right]}.
\end{equation}
Moreover, we compute the spatial domains' scale for $C(l)=0.15$, where $l$ is the characteristic length.

We calculated the spatial autocorrelation function 
in terms of the radial coordinate $r$ for three cases: absence of grouping behaviour ($\varphi=1.0$), aggregation triggered when the neighbourhood is, at least, $10\%$ hostile ($\varphi=0.1$), and non-adaptive aggregation ($\varphi=0.0$). The outcomes were obtained by running sets of $100$ simulations with different random initial conditions in lattices with $500^2$ grid sites for a time span of $5000$ generations. To calculate the autocorrelation function, we used the spatial configuration at the end of the simulation $(t=5000$ generations). Because organisms of every species can perform the locally adaptive congregation, the autocorrelation function is the same irrespective of the species; thus, we used the data from species $1$. In all simulations, we considered the interactions probabilities $s=r=0.25$ and $m=0.5$; the perception radius was set to $R=3$.

\begin{figure}[t]
	\centering
	\includegraphics*[width=8.5cm]{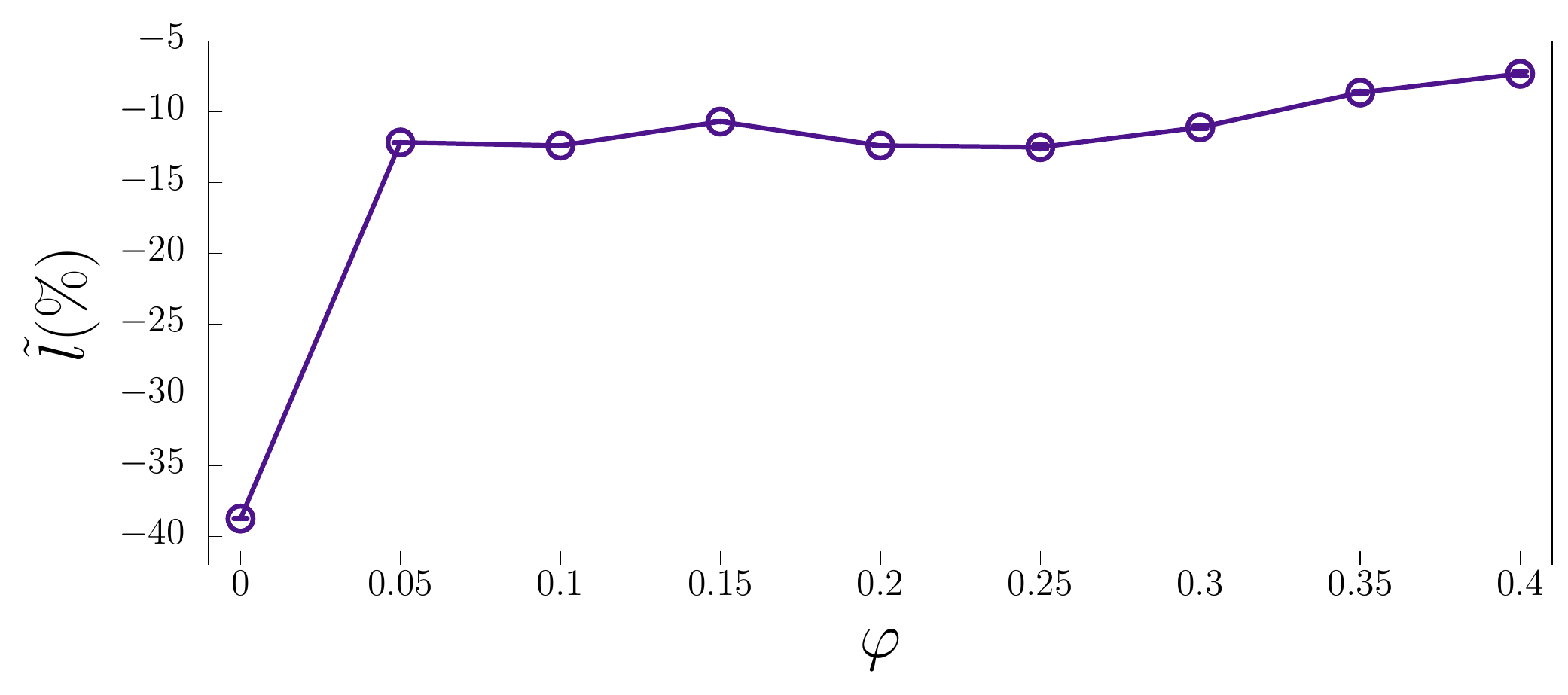}
\caption{The relative change in the characteristic length scale of the typical single-species spatial domain as a function of the threshold used to trigger the gregarious movement compared with the standard model. The simulations ran in lattices with $500^2$ grid sites, running until $5000$ generations for $r=s=0.25$ and $m=0.5$; the perception radius is $R=3$.
The outcomes were averaged from sets of $100$ simulations starting from different initial conditions; the error bars show the standard deviation. We assumed the probabilities $r=s=0.25$ and $m=0.5$.
} 
 \label{fig6}
\end{figure}

The brown, orange, and grey lines in Figure \ref{fig5} show
$C$ as a function of the radial coordinate $r$ 
for $\varphi=0.0$, $\varphi=0.1$, and $\varphi=0.0$, respectively; the error bars indicate the standard deviation. 
The horizontal dashed black line indicates the threshold used to calculate the length scale: $C(l)\, =\, 0.15$. The results confirm that once organisms move gregariously, the average size of the spatial domains inhabited by a single species decreases. 

Figure \ref{fig6} shows the relative variation of the characteristic length scale $\tilde{l}$, defined as $\tilde{l}=(l-l_0)/l_0$, where $l_0$ is the value in the absence of the adaptive aggregation ($\varphi=1.0$). We repeated the set of $100$ simulations - starting from different initial conditions - for
$0 \leq \varphi \leq 0.4$, with intervals of $\delta \varphi =0.05$. The error bars show the standard deviation; the parameters are the same used in the simulations in Fig.~\ref{fig5}. The outcomes show that the average group size decreases compared to the standard model, with the reduction becoming significant for $\varphi = 0.0$. This happens because all individuals group themselves, independently of what is happening in their surroundings, as we observed in Fig.~\ref{fig2d}.

\section{The role of the locally adaptive aggregation in the organisms' death risk}
\label{sec5}

We now investigate the effects of locally adaptive grouping to reduce the organisms' death risk. For this purpose, we introduce the death risk, which is calculated as follows: i) it is counted as the total number of individuals of species $i$ at the beginning of each generation; ii) the number of organisms of species $i$ killed by individuals of species $i-1$ during the generation is computed; iii) the death risk, $\zeta$ is defined as the ratio between the number of eliminated organisms and the amount at the beginning of each generation. Due to the symmetry of the rock-paper-scissors game rules, the average death risk is the same for individuals of every species; thus, we choose the results for species $1$ to represent the individuals' death risk.

\begin{figure}
	\centering
	\includegraphics*[width=8.5cm]{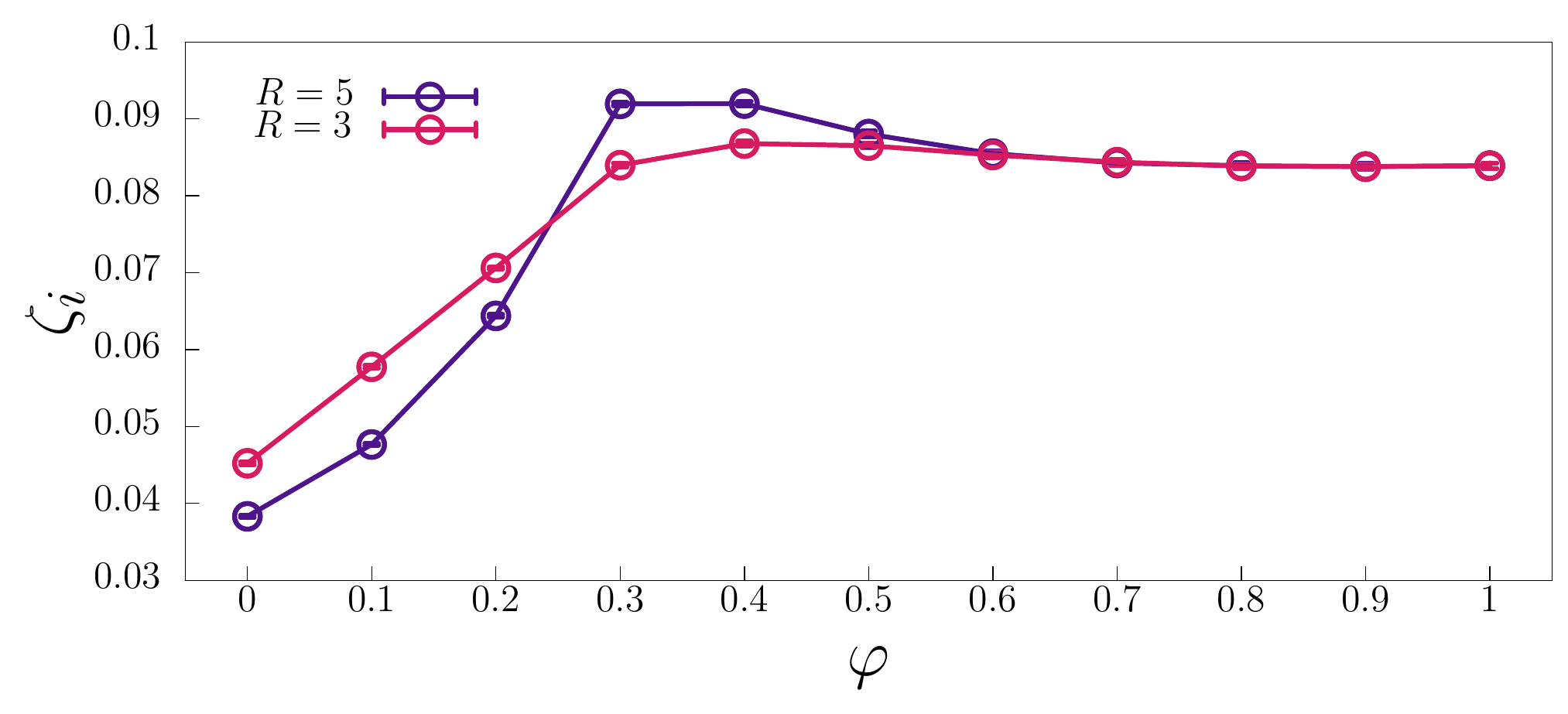}
\caption{Organisms' death risk in terms of the aggregation trigger.
The simulations were performed in lattices with $500^2$ grid sites, running for a timespan of $5000$ generations.
The red and purple lines show the outcomes for organisms with perception radius $R=3$ and $R=5$, respectively. The results were averaged from sets of $100$ simulations starting from different initial conditions; the standard deviation is depicted by error bars. The interaction probabilities are $s=r=0.25$ and $m=0.5$.}
 \label{fig7}
\end{figure}

\begin{figure}[t]
	\centering
	\includegraphics*[width=8.5cm]{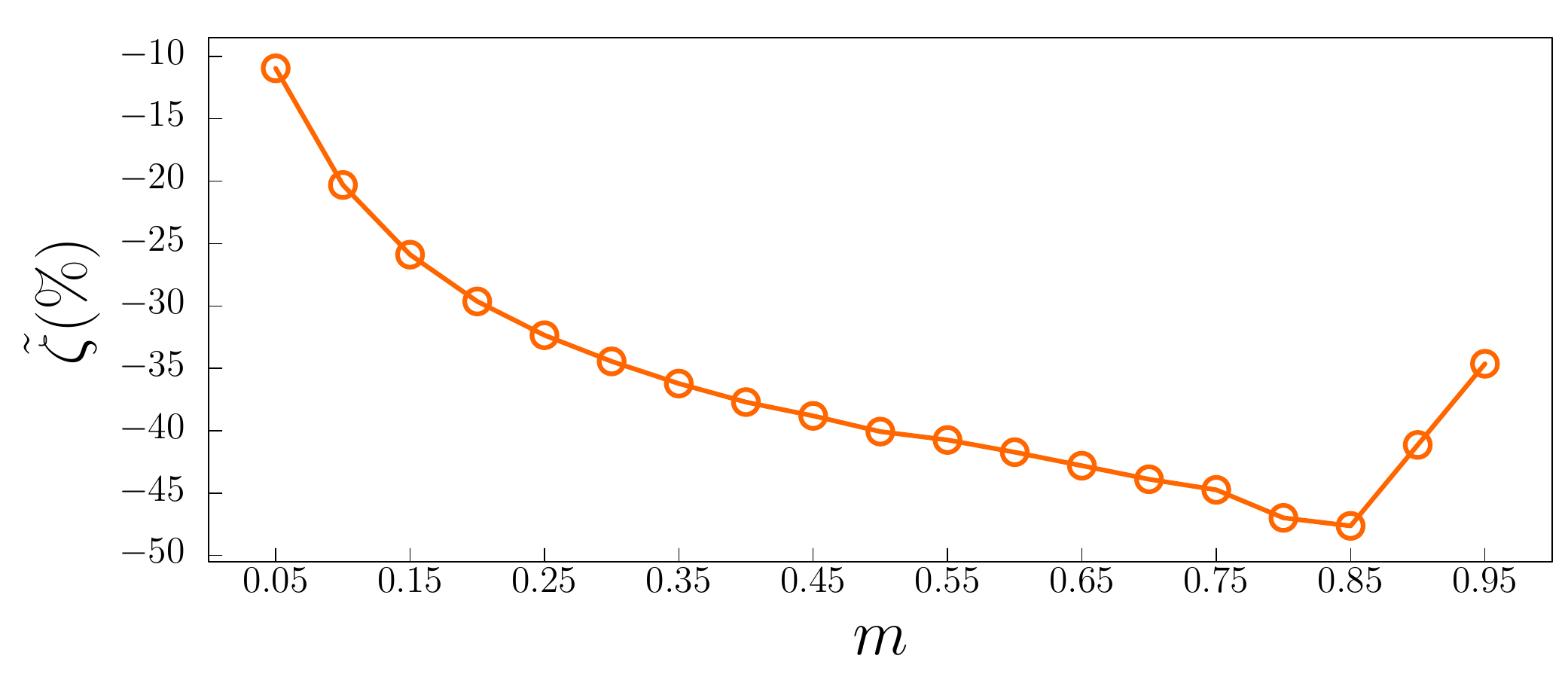}
\caption{Relative change in the individuals' death risk in terms of the mobility probability in simulations running in lattices with $500^2$ grid sites, running for a timespan of $5000$ generations.
We averaged the outcomes sets of $100$ simulations starting from different initial conditions; the standard deviation is shown by error bars. The perception radius is $R=3$; the interaction probabilities are to $s=r=(1-m)/2)$.}
 \label{fig8}
\end{figure}

\subsection{The influence of the aggregation trigger}

First, we study the influence of the aggregation trigger $\varphi$ in the relative decrease of the individuals' death risk by running sets of $100$ simulations starting from different initial conditions for $0 \leq \varphi \leq 1.0$ in intervals of $\delta \varphi=0.1$. This experiment was conducted for two values of perception radius: $R=3$ and $R=5$; the interaction probabilities are $s=r=0.25$ and $m=0.5$. To guarantee the quality of the results, we remove the data from the initial pattern formation stage, thus calculating the average organisms' death risk in the second half of each realisation. 

The purple and red lines in Figure \ref{fig7} show the organisms' death risk in terms of the aggregation trigger for $R=3$ and $R=5$, respectively; the standard deviation is shown by error bars. The outcomes reveal that for $\varphi \geq 0.6$, the locally adaptive strategy is ineffective in reducing the organisms' death risk compared with the standard model ($\varphi=1.0$). This happens because  most of organism of species $i$ whose neighbourhood contains $60\%$ or more of organisms of species $i-1$ is far from the spatial domain dominated by their conspecifics; thus, grouping may not be possible to be executed before the individual being eliminated by enemies.

Our findings show that the locally adaptive aggregation jeopardises the organisms' safety for intermediate values of $\varphi$. As shown in Fig.~\ref{fig7}, for $R=3$, the organisms' death risk increases for $0.4 \leq \varphi <0.6$, while for $R=5$, $\zeta$ increases for $0.4 \leq \varphi <0.3$. Therefore, the adaptive is beneficial only if the threshold assumed to move gregariously is in the interval $0 \leq \varphi <0.4$ for $R=3$ and $0 \leq \varphi <0.3$ for $R=5$, with the relative reduction of $\zeta$ increasing as the $\varphi$ is lowered.

The results in Fig.~\ref{fig7} show how the complexity of the spatial interactions is influenced by the organism's ability to make an accurate decision, triggering the adaptive tactic correctly. 
Our findings show that if organisms can perceive further distances, they can more easily: i) identify the presence of invading enemies beyond the border of their territory; ii)
 distinguish the direction with more conspecifics in case of need to move gregariously. Because of this, the relative variation in the organisms' death risk is more accentuated for $R=5$ than for $R=3$ in Fig.~\ref{fig7}.

\subsection{The interference of organisms' mobility}

The locally adaptive grouping is profitable for the organisms because of the death risk reduction, as shown in Fig.~\ref{fig7} for $m=0.5$. Now, we repeated the simulations to explore how the benefits of the locally adaptive aggregation depend on the organism's mobility. For this purpose, we ran sets of $100$ realisations starting from different initial conditions for $0.05 \leq m \leq 0.95$, in intervals of $\delta m = 0.05$. The selection and reproduction probabilities are set to $s=r=(1-m)/2$; the perception radius is $R=3$, and the aggregation trigger is $\varphi=0.05$. We implemented the simulations in lattices with $500^2$ grid sites, running until $5000$ generations. 

Figure \ref{fig8} shows the relative change of the organisms' death risk: $\tilde{\zeta}=(\zeta-\zeta_0)/\zeta_0$, where $\zeta_0$ is the death risk in the absence of grouping behaviour ($\varphi=1.0$).
For $0.05 \leq m \leq 0.085$, 
the relative reduction in the organisms' death risk is more significant for individuals that explore greater 
fractions of the lattice per time unit \cite{mobilia2}. This happens because high-mobile individuals are more vulnerable to being eliminated by enemies in the cyclic game, thus, benefitting more from the self-preservation movement strategy. 
However, if $m > 0.085$, the relative variation in $\zeta$ decreases because the selection probability becomes very low, becoming the effect less significant.
\begin{figure}[h]
\centering
       \begin{subfigure}{.48\textwidth}
        \centering
        \includegraphics[width=85mm]{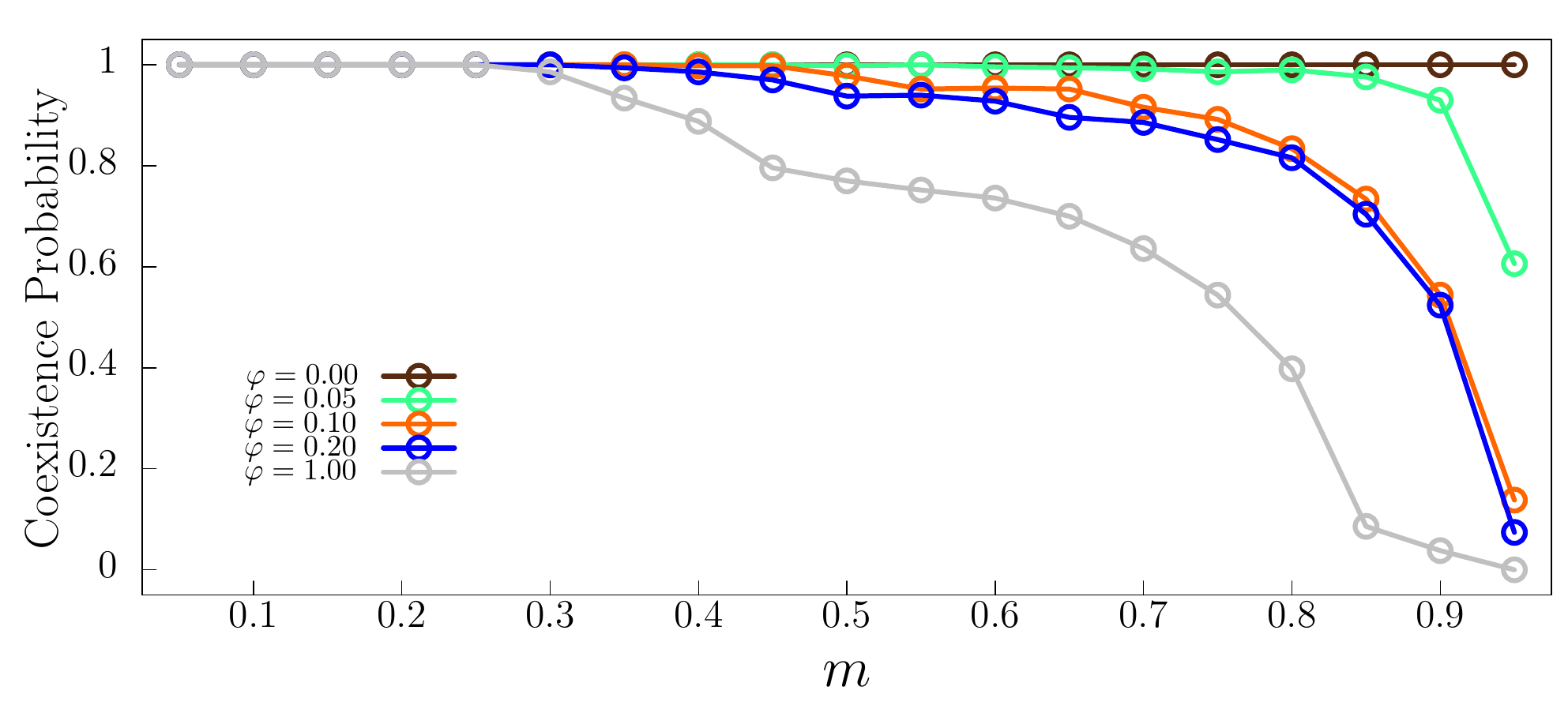}
        \caption{}\label{fig9a}
    \end{subfigure}\\
           \begin{subfigure}{.48\textwidth}
        \centering
        \includegraphics[width=85mm]{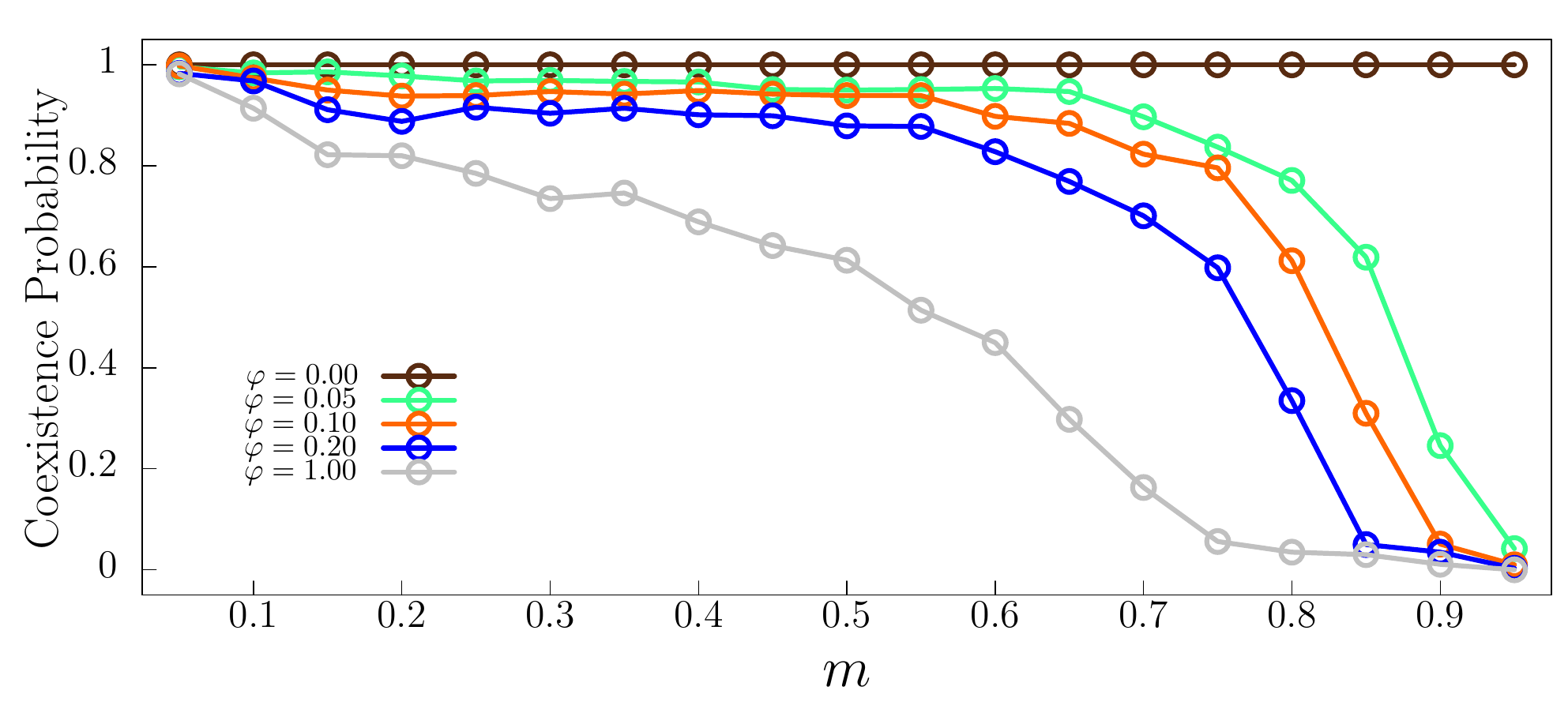}
        \caption{}\label{fig9b}
    \end{subfigure}\\
           \begin{subfigure}{.48\textwidth}
        \centering
        \includegraphics[width=85mm]{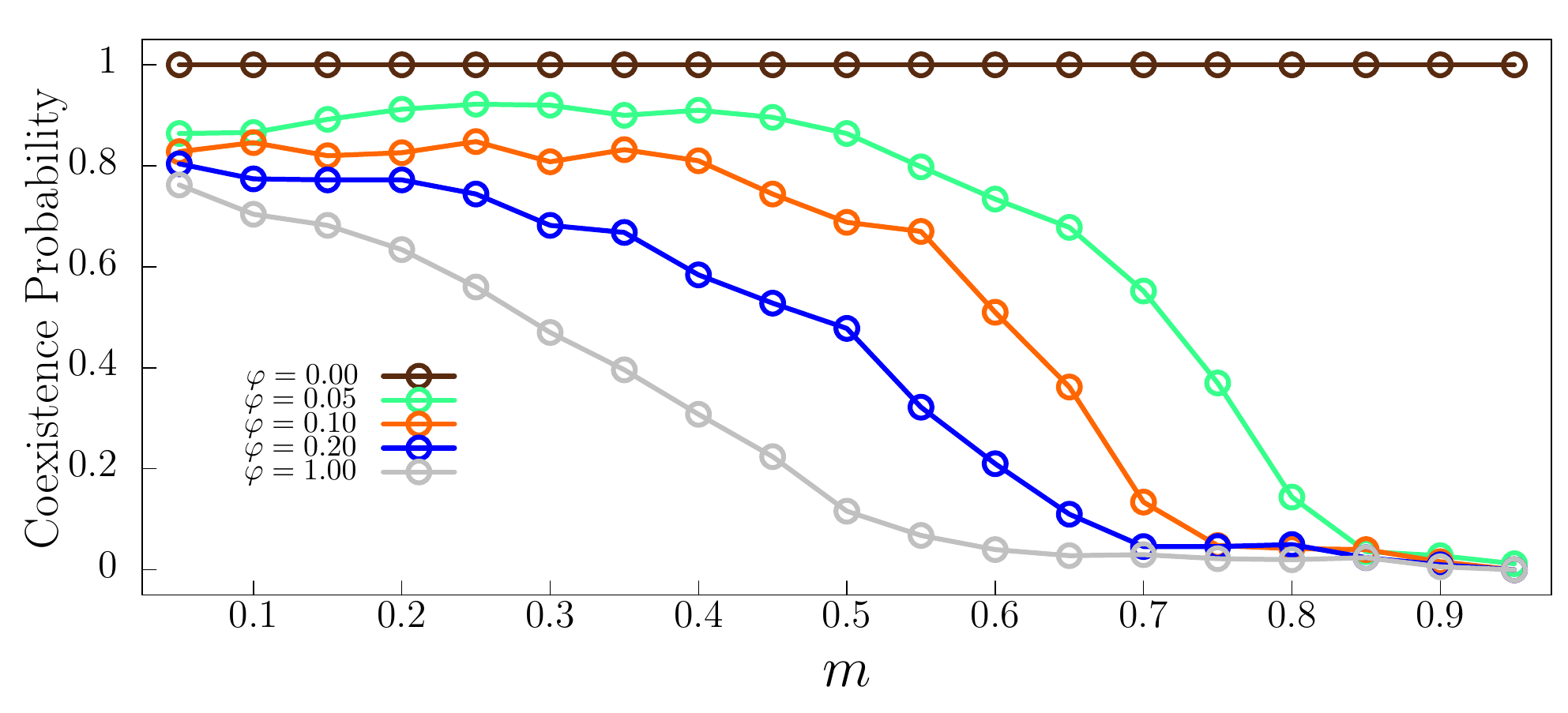}
        \caption{}\label{fig9c}
    \end{subfigure}
\caption{Coexistence probability as a function of the mobility probability for the generalised rock-paper-scissors game with organisms' locally adaptive aggregation.
Figures \ref{fig9a}, \ref{fig9b}, and \ref{fig9c} show the outcomes for the cyclic model with three, five, and seven species, respectively. The results were obtained by running $1000$ simulations in lattices with $100^2$ grid points running until $10000$ generations for $R=3$ and $s=r=(1-m)/2$.
The brown, green, orange, blue, and grey lines depict the results for $\varphi=0.0$, $\varphi=0.05$, $\varphi=0.1$, $\varphi=0.2$, and $\varphi=1.0$, respectively.}
\label{fig9}
\end{figure}
\section{Coexistence Probability}
\label{sec6}
Now, we focus on the impact of locally adaptive flocking on biodiversity in cyclic games. In this study, we ran sets of 
 $1000$ simulations in lattices with $100^2$ grid points for $ 0.05\,<\,m\,<\,0.95$ in intervals of $ \delta\, m\, =\,0.05$; selection and reproduction probabilities were set to $s=r\,=\,(1-m)/2$. For each set of simulations, each realisation began from different random initial conditions, running until $10000$ generations. If at least one species is extinguished before the simulation ends,  biodiversity is lost. Thus, the coexistence probability is the fraction of the simulations where all species are present at the end.
We extended the investigation to quantify the impact of locally adaptive aggregation in more complex systems by simulating the generalised rock-paper-scissors models with five and seven species. 
Figures \ref{fig9a}, \ref{fig9b} and \ref{fig9c} depict the coexistence probability for $\varphi=0.0$ (brown line), $\varphi=0.05$ (green line), $\varphi=0.1$ (orange line), $\varphi=0.2$ (blue line), and $\varphi=1.0$ (grey line) for the models with $N=3$, $N=5$, and $N=7$ species, respectively. 

Overall, species biodiversity is more threatened for systems with highly mobile individuals, independent of the number of species in the cyclic game. The outcomes also show the benefits of the locally adaptive aggregation for biodiversity: the lower the aggregation trigger, the higher is the coexistence probability. This conclusion holds independently of the number of species in the cyclic game
Furthermore, the outcomes show that the more complex the system is, the more favourable it is for biodiversity loss. By comparing the same color lines in Fig.~\ref{fig9a}, \ref{fig9b} and \ref{fig9c}, one observes that the coexistence probability is lower for the system with $N=9$ species, independently of the organisms' mobility.  
Finally, we observe that all simulations resulted in coexistence when individuals agglomerate with their conspecifics irrespective of the local densities of enemies. 
According to the brown lines in Figs.~\ref{fig9a}, \ref{fig9b} and \ref{fig9c}.

\section{Comments and Conclusions}
\label{sec7}

Aggregation behaviour is found in many systems where organisms adapt their movement, grouping with their conspecifics when in death risk. We investigate cyclic models described by the 
rock-paper-scissors game rules, where individuals can scan their environment and adapt their movement to environmental cues. In our stochastic simulation, each organism freely explores the
territory without precaution if there is no nearby enemy but prevents damage from enemy attack moving gregarious to join the biggest group of conspecific in the neighbourhood. To execute the locally  adaptive grouping, each individual scans their vicinity, thus triggering the gregarious movement if the local density of enemies reaches a prefixed threshold. Running a series of simulations, we investigate the role of adaptive aggregation in transforming the organisms' spatial organisation. The results show that the characteristic length scale of the spatial domains occupied by organisms of a single species is not accentuated if the threshold is not inferior to $10\%$. Otherwise, the typical group size decreases significantly, being minimal in the case of organisms flock even when not under death risk pressure.

We discover that 
the gregarious movement does not interfere with organisms' safety if the grouping is only triggered when more than $70\%$ neighbourhood is occupied by enemies. Counterintuitively, if the self-preservation movement tactic is calibrated to be triggered if between $30\%$ and $60\%$ neighbours are enemies, the individuals' death risk increases instead of benefiting the organisms. Our outcomes show that the behavioural strategy is profitable only if each organism aggregates with conspecifics when detecting the fraction of opponents in the vicinity using a threshold inferior to $30\%$. In addition, we find that if organisms can perceive further distances, they can accurately scan and interpret the signals from the neighbourhood, increasing the effects of the adaptive aggregation on the death risk. Moreover, we study the impact of mobility on the benefits of adaptive congregation considering low, intermediate and high-mobile individuals. Our simulations provided evidence that locally adapting their movement to aggregate when under death risk is more advantageous as the more mobile the organisms, provided that the individuals' mobility is not superior to $85\%$; otherwise, the relative death risk reduction diminishes as the mobility grows.

Finally, we study the influence of locally adaptive aggregation on biodiversity maintenance. Our findings show that the coexistence probability increases independently of the organism's mobility, being maximal in the case of non-adaptive grouping, where the gregarious movement is executed even when there is no local death risk for the individual. This result holds for more complex systems where an arbitrary odd number of species participate in the cyclic game. Extending our algorithm to implement the generalised rock-paper-scissors model with five and seven species, we confirm that the gregarious movement promotes biodiversity, being more beneficial for low adaptive aggregation triggers.
Our discoveries may be helpful to ecologists in understanding 
systems where organisms' self-defence behaviour adapts to local environmental cues.
Our results may also clarify the role of the local phenomena in complex systems in other areas of nonlinear science.

\section*{Acknowledgments}
We thank CNPq, ECT, Fapern, and IBED for financial and technical support.

\bibliographystyle{elsarticle-num}
\bibliography{ref}

\end{document}